\documentclass[twocolumn]{article}

\usepackage{color}
\usepackage[utf8]{inputenc}
\usepackage{authblk}
\usepackage{url}
\usepackage{hyperref}
\hypersetup{colorlinks=false}
\usepackage{fullpage}
\usepackage{graphicx}
\usepackage{abstract}

\title{A Nonzero Gap Two-Dimensional Carbon Allotrope from Porous Graphene}

\author[1]{G. Brunetto}
\author[1,*]{P. A. S. Autreto}
\author[1]{L. D. Machado}
\author[1]{B. I. Santos}
\author[2]{R. P. B. dos Santos}
\author[1]{D. S. Galv\~ao}

\date{\vspace{-5ex}}

\affil[1]{Instituto de F\'{\i}sica ``Gleb Wataghin", Universidade Estadual de Campinas, 13083-970 Campinas SP, Brazil}
\affil[2]{Departamento de F\'{\i}sica, IGCE, UNESP, 13506-900 Rio Claro SP, Brazil}
\affil[*]{To whom correspondence should be addressed. Email: \url{autretos@ifi.unicamp.br}}

\begin{document}
\twocolumn[
\maketitle

\begin{onecolabstract}
Graphene is considered one of the most promising materials for future electronic. However, in its
pristine form graphene is a gapless material, which imposes limitations to its use in some
electronic applications. In order to solve this problem many approaches have been tried, such as, physical and chemical functionalizations. These processes compromise some of the desirable graphene properties. In this work, based on \textit{ab initio} quantum molecular dynamics, we showed that a two-dimensional carbon allotrope, named biphenylene carbon (BPC) can be obtained from selective dehydrogenation of porous graphene. BPC presents a nonzero bandgap and well-delocalized frontier orbitals. Synthetic routes to BPC are also addressed.
\end{onecolabstract} 
]


\section{Introduction}

In the last decades the successive discoveries of new carbon-based materials have opened a new era in materials science. Examples of these discoveries are fullerenes \cite{fullerenes}, carbon nanotubes \cite{iijima}, and more recently, graphene \cite{geim1,geim2}.

Graphene is a two-dimensional array of hexagonal units of $sp^2$ bonded carbon atoms (Fig. 1a). Graphene presents very unusual and interesting electronic and mechanical properties \cite{geim1,geim2}. Because of these special properties graphene is considered one of the most promising materials for future electronics. However, in its pristine form graphene is a gapless semiconductor, as shown in Fig. 2a. This poses serious limitations  to its use in some electronic applications, such as, some kind of transistors \cite{withers}.

Many approaches have been tried in order to create a gap in graphene-like materials. The most common strategies use chemical and physical methods, such as, oxidation \cite{ruoff}, hydrogenation \cite{sluiter2003,sofo,elias,flores} and fluorination \cite{humberto, geim3, leenaerts2010}. However, the controlled synthesis of large structures and/or at large scale have been proved to be difficult. More important, the desirable electronic graphene properties are partially compromised in such approaches.

\begin{figure}[tb]
\includegraphics[width=\columnwidth]{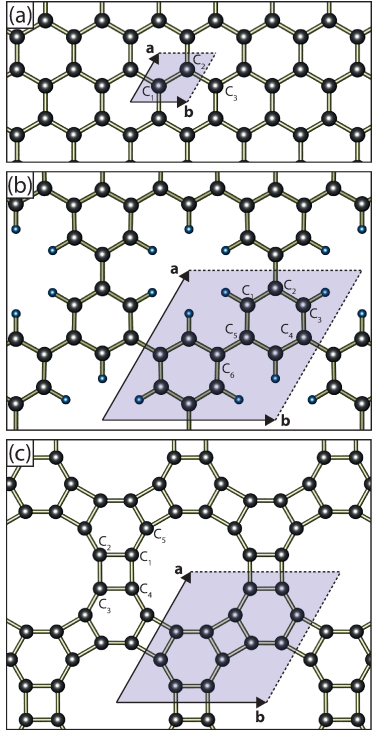}
\caption{(Color online)Structural models considered in the present work. 
(a) Graphene, with two atoms in the unit cell, (b) porous graphene with 18 atoms in the unit cell
and (c) BPC with 12 atoms in the unit cell. Unit cell is highlighted in each case.}
\label{estruturas}
\end{figure}

Another approach has been trying to obtain intrinsically hydrogenated structures, such as, the so-called porous graphene (PG) (Fig. 1b) \cite{bieri,blankerbourg,du,jiang,li}, whose synthesis has been recently achieved \cite{bieri}. But again, the obtained structures present some of the same problems of chemically/physically functionalized graphene, such as, excessive large bandgap value and flat (low mobility) electronic bands (Fig. 2b).

An ideal structure would be an allotrope carbon form with an intrinsic good bandgap value and electronic bands with good dispersion (electron delocalization and high charge mobility). In theory, structures satisfying these conditions do exist, as the so-called biphenylene carbon (BPC) (Figs. 1c and 2c) \cite{bpc,bpc2}. Molecular fragments in linear, zigzag and several other forms, have already been synthesized \cite{bpc3}. They are considered as potential precursors for fullerenes, bowls, cyclophanes, etc. \cite{bpc,bpc2,bpc3}. However, the synthesis of large BPC fragments remains elusive.

In this work, based on \textit{ab initio} molecular dynamics simulations, we show that selective dehydrogenation of porous graphene leads to the spontaneous interconversion to BPC structures. Possible synthetic route approaches to achieve this interconversion are also discussed. That this interconversion was possible was found out by exploratory investigations, as discussed below.

\begin{center}
\begin{figure}[!t]
\includegraphics[width=\columnwidth]{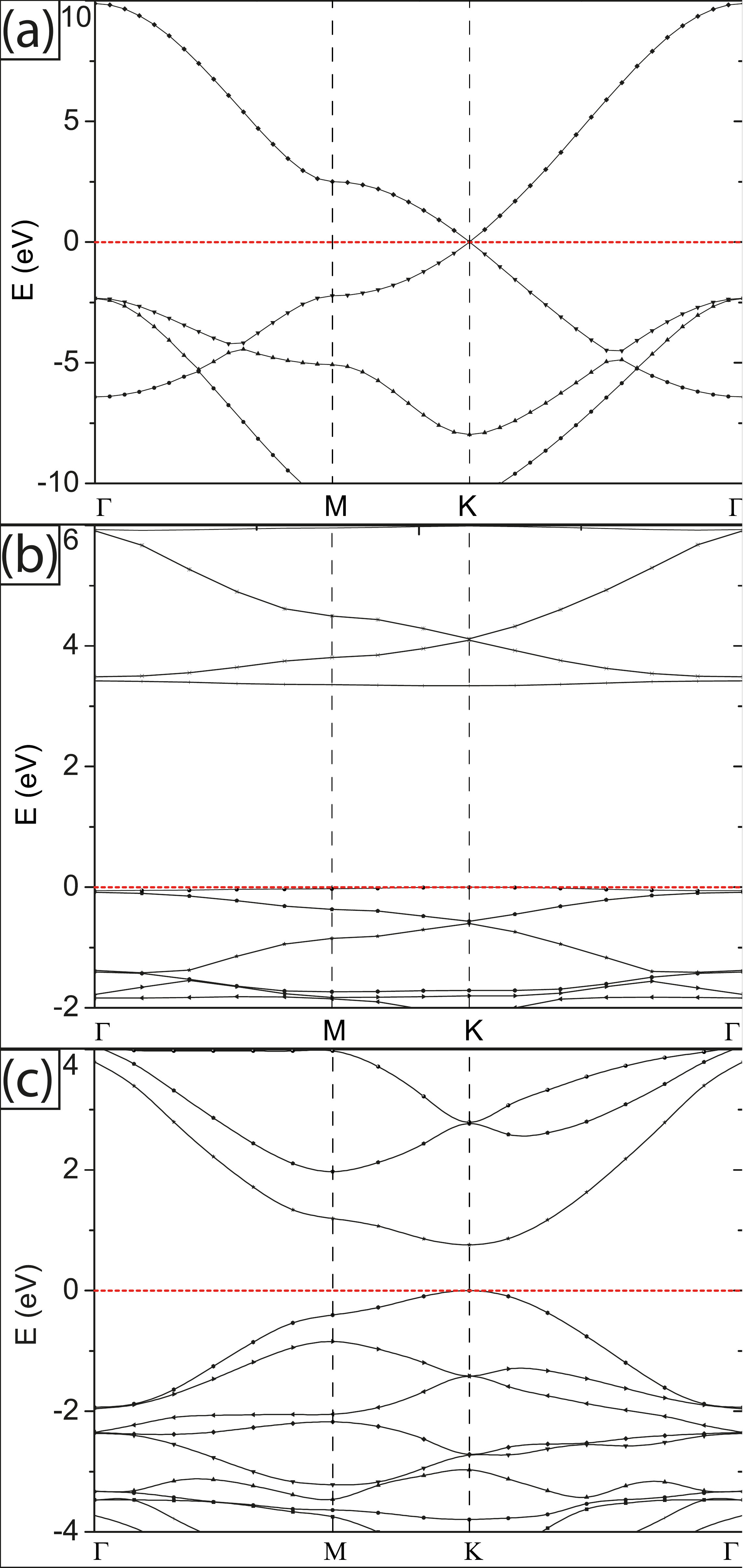}
\caption{(color online)Band structure (in eV) for the (a) graphene, 
(b) porous graphene and (c) BPC. Dashed red lines indicate the Fermi level.
Please notice that DFTB+ convention to locate the Fermi level is at the value of the highest occupied state.}
\label{bandas}
\end{figure}
\end{center}

\section{Methodology}

The geometric and electronic aspects of the structures shown in Fig. 1, as well as, the interconversion processes from PG to BPC, were investigated in the framework of density functional theory (DFT). Exchange and correlation terms were considered within the generalized gradient approximation (GGA) with a BLYP functional \cite{gga}, with a double numerical plus polarization basis set, as implemented in DMol$^3$ code \cite{dmol,dmol2}. In all the calculations a non-relativistic and all-electron treatment was used. The parameter criteria for the tolerances of energy, force, displacement and SCF convergence criteria were $2.72\times 10^{-4}$ eV, $5.44\times 10^{-2}$ eV $\cdot$ \AA$^{-1}$, $5.0\times 10^{-3}$ \AA\, and $1.0\times 10^{-6}$, respectively. 

It is well-known that although DFT methods can reliably describe the geometrical features, in general some electronic properties, as the bandgap values, are underestimated \cite{dft}. For these reasons, for the analysis of the electronic band structure calculations we used a DFT based tight-binding method (DFTB+) \cite{dftb1,dftb2}, which has been proved to reliably describe the electronic properties of carbon-based materials. The DFTB+ calculations were carried out using the optimized geometries from the DMol$^3$ calculations.
The DFTB+ non-diagonal matrix (overlap matrix and Hamilton matrix) elements are calculated in a two-center approximation. They are distant dependent and considered up to about 10 atomic units.

Initially, we carried out DMol$^3$ fully (unit cell parameters and atomic positions were allowed to vary) geometric optimizations. The most relevant geometrical data are displayed in Tables I and II. See also supplementary materials.

\begin{table}
\caption{DMol$^3$ geometrical data. Labels according to Fig. \ref{estruturas}}
\begin{center} {\bf GRAPHENE } \end{center}
\begin{tabular}{lcc|lc}
\multicolumn{2}{c}{Distances (\AA)} & & \multicolumn{2}{c}{Angles ($^o$)} \\
\hline
C$_1$-C$_2$ & 1.39 & & C$_1$-C$_2$-C$_3$ & 119.96 \\
\end{tabular}
\begin{center} {\bf POROUS GRAPHENE} \end{center}
\begin{tabular}{lcc|lc}
\multicolumn{2}{c}{Distances (\AA)} & & \multicolumn{2}{c}{Angles ($^o$)} \\
\hline
C$_1$-C$_2$ & 1.41 & & C$_1$-C$_5$-C$_6$ & 121.56 \\
C$_3$-C$_4$ & 1.41 & & C$_2$-C$_1$-C$_5$ & 122.68 \\
C$_5$-C$_6$ & 1.50 & & C$_1$-C$_2$-C$_3$ & 117.77 \\
\end{tabular}
\begin{center} {\bf BPC} \end{center}
\begin{tabular}{lcc|lc}
\multicolumn{2}{c}{Distances (\AA)} & & \multicolumn{2}{c}{Angles ($^o$)} \\
\hline
C$_1$-C$_2$ & 1.476 & & C$_1$-C$_2$-C$_3$ & 89.826 \\
C$_1$-C$_4$ & 1.485 & & C$_2$-C$_3$-C$_4$ & 90.174 \\
C$_1$-C$_5$ & 1.365 & & C$_2$-C$_1$-C$_5$ & 119.846 \\
\end{tabular}
\label{tab:tabela1}
\end{table}

\section{Results}

Our results show that the PG optimized geometry is characterized by a $C222$($D_{2}^{6}$) symmetry group. The hexagonal geometry is preserved, with bond-length in the rings and intra-rings of 1.40 and 1.50 \AA, respectively. The obtained geometry is in good agreement with the experimental data reported \cite{bieri}, theory and experiment estimate the lattice parameter to be about 7.4 \AA. The obtained bandgap value of 3.3 eV is also in good agreement with previous theoretical calculations (3.2 eV) \cite{du}.

For the BPC layer, a stable conformation was obtained and satisfying the topological conditions of the theoretically proposed structures \cite{bpc,bpc2}. In relation to PG, there is a significant lattice parameter contraction, from 7.5 to 6.8 \AA, respectively (Table II). The bond-lengths in the hexagons preserve the pattern of alternating double and single bonds (1.48 and 1.36 \AA, respectively), while the square structure (cyclobutadiene) is almost a perfect square (1.48 \AA) (see Fig. 1 and Table I).

\begin{table}
\caption{Lattice parameters for graphene, porous graphene and biphenylene carbon (BPC), respectively.}
\begin{tabular}{lccc}
 & {\bf a} (\AA) & {\bf b} (\AA)  \\ 
\hline
Graphene 		& 2.40 & 2.40  \\
Porous Graphene & 7.52 & 7.53  \\
BPC 			& 6.78 & 6.69  \\
\end{tabular}
\label{tab:tabela2}
\end{table}

In Fig. 2 we present the band structure results obtained from DFTB+ calculations. As expected, graphene presents a zero bandgap value, while PG and BPC have values of 3.3 and 0.8 eV, respectively. 
However bandgap values and good dispersion of the frontier bands are not warranty of good conductors. Besides these aspects another important characteristic which plays an important role in defining the electronic (conductivity) mobility of the material is the degree of electronic delocalization of the frontier crystalline orbitals. 
The electronic analysis we carried out here for ideal BPC structures are for neutral forms (no free carriers). In order to create these carriers is necessary, as usual, to dope the material. In ref [16] the doping of porous graphene has been addressed. The same principles can be used to dope BPC structures.

\begin{figure}[tb]
\includegraphics[width=\columnwidth]{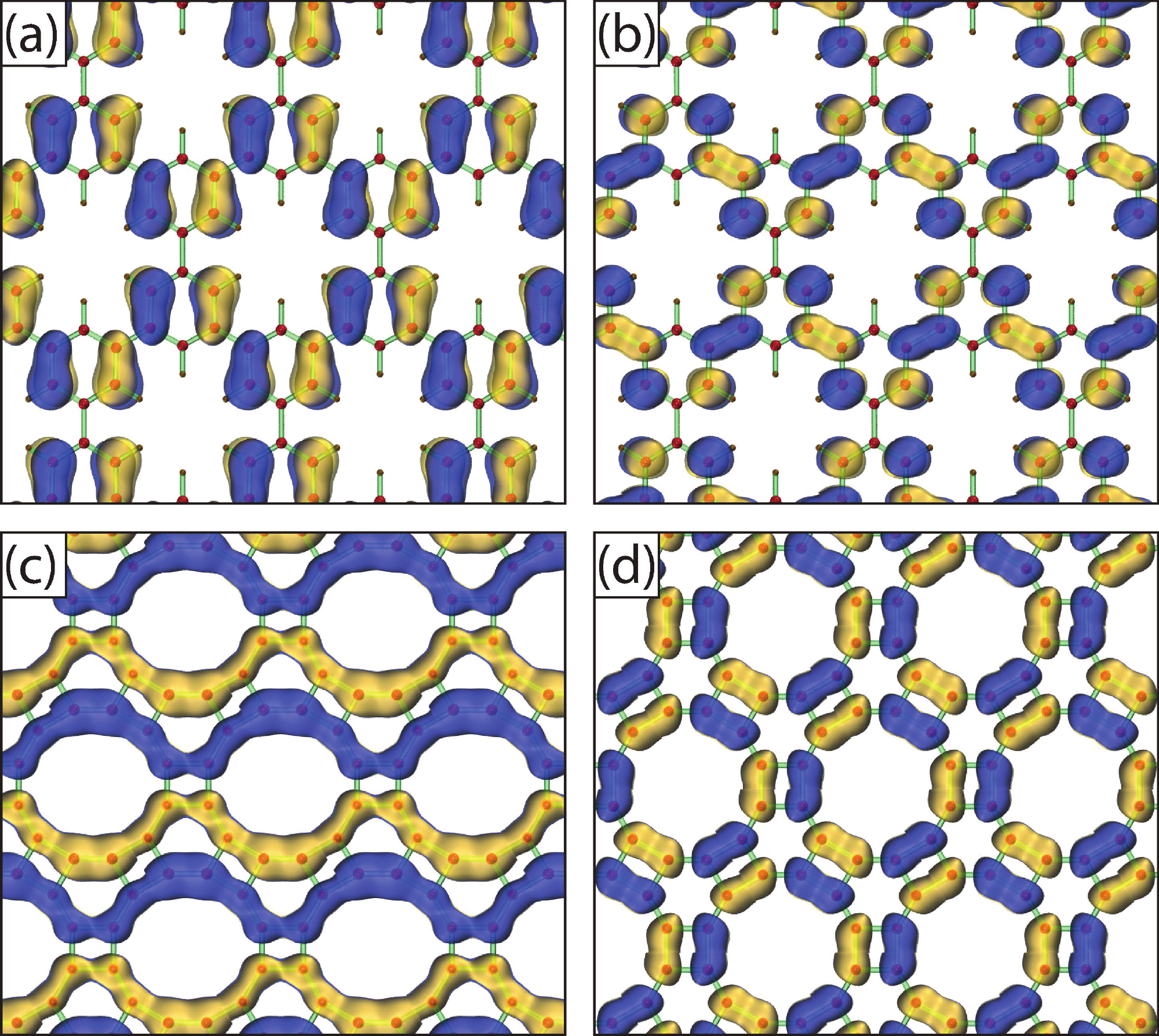}
\caption{(Color online)DFTB+ frontier orbitals for hydrogenated and dehydrogenated
porous graphene.
{\bf (a)} and {\bf (b)} HOCO and LUCO for porous graphene.
{\bf (c)} and {\bf (d)} HOCO and LUCO for dehydrogenated porous graphene.}
\label{orbitals}
\end{figure}

In Fig. 3 we present the frontier orbitals HOCO (highest occupied crystalline orbital) and LUCO (lowest unoccupied crystalline orbital) for PG and BPC. As we can see from Fig. 3 BPC presents more delocalized orbitals than PG. Particularly interesting is the BPC HOCO, which is well delocalized over the whole network, suggestive of a good conductor structure. Thus, in principle, BPC presents itself as an ideal structure for many electronic applications; good intrinsic bandgap value, bands with good dispersion and delocalized frontier orbitals.
Also, we calculated the values of the effective masses $(m^*/m_e=0.26$ and $m^*/m_h=0.33$, for the conduction and valence effective masses, respectively). These values are consistent with the expected to a good conductor.
However, as mentioned before, the synthesis of large BPC fragments from using chemical methods remains elusive. Through exploratory investigations we found out that selective hydrogen removal from PG leads to a spontaneous interconversion to BPC.

\begin{figure}[ht!]
\includegraphics[width=\columnwidth]{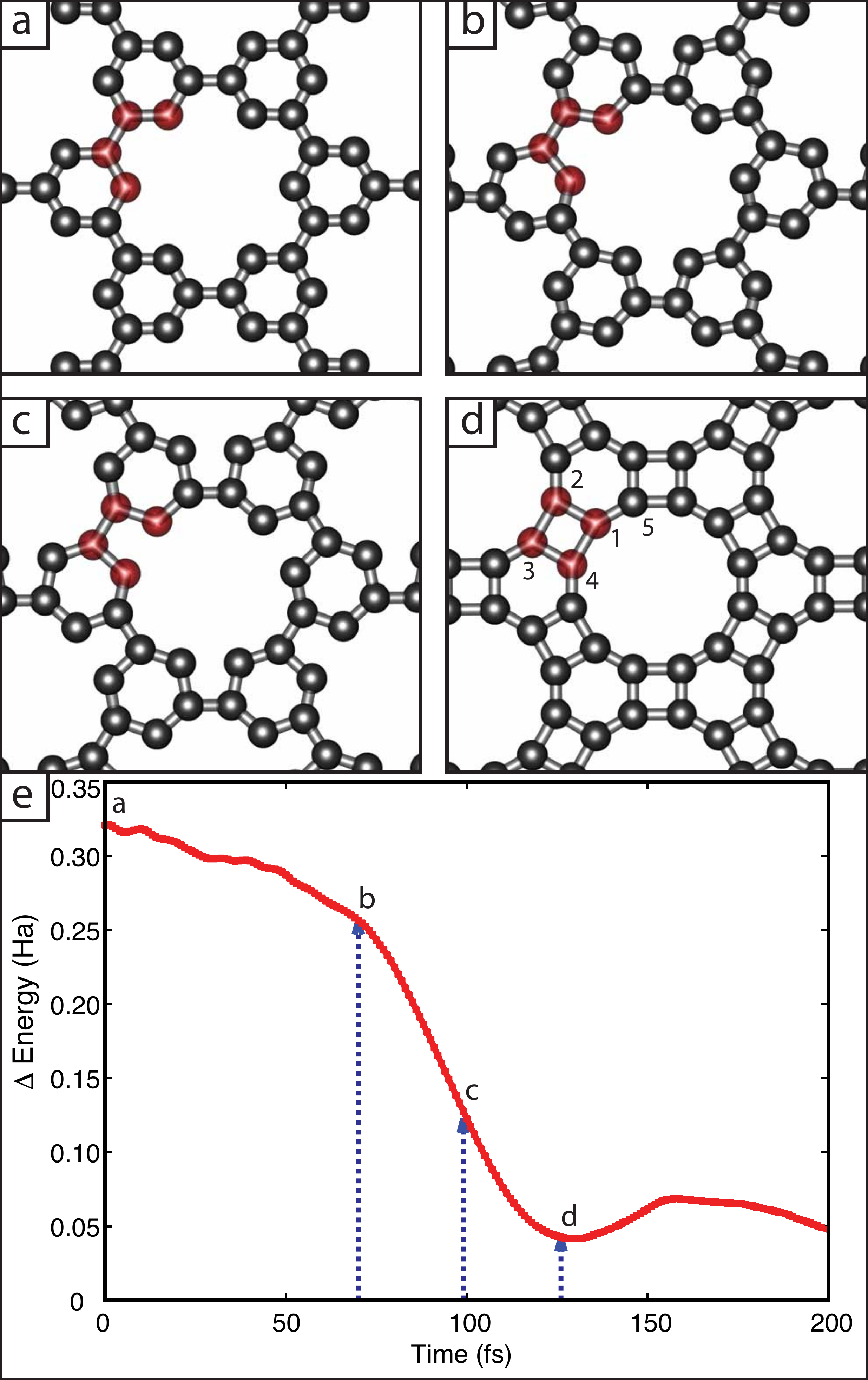}
\caption{DMol$^3$ molecular dynamics snapshots and total energy values as a function of time. 
{\bf (a)} Completely dehydrogenated porous graphene on the
initial stage. The highlighted carbons in red lead to the 
formation of a cyclobutadiene. {\bf (b)} and {\bf (c)} Intermediate 
stages. Notice that adjacent rings rotate in opposite directions.
{\bf (d)} Final stage after a complete rotation of $30^o$ in
each ring. {\bf (e)} Time evolution for the total energy during the
QMD simulation. Letter labels represent the energy of each snapshot.
}
\label{dinamica}
\end{figure}

We have carried out \textit{ab initio} quantum molecular dynamics (DMol$^3$), NVT ensemble, at different temperatures: 0 (just geometric optimizations), 300 and 600 K. We started from the optimized PG geometry, then we removed the hydrogen atoms and let the system to freely evolve in time (lattice parameter values and atomic positions free to vary). For all the investigated cases we observed a spontaneous interconversion from dehydrogenated PG to BPC. The obtained BPC structures are thermally stable (at least up to 600 K).

In Fig. 4 we present results for the calculations carried out at 300 K. In Figs. 4a-4d we show snapshots from the molecular dynamics simulations at successive time steps. We can see that the dehydrogenated PG undergoes structural rearrangements, mainly ring rotations (about 30 degrees) coupled to a lattice parameter reduction leading to the formation of a cyclobutadiene motif (highlighted atoms in Figs. 4a-4d), and consequently to the BPC formation. The whole process can be better visualized in the video01 of the supplementary materials. 

In Fig. 4e we present the total energy values as a function of time of simulation. As we can see, starting from dehydrogenated PG the system continuously evolve to more stable configurations reaching a well-defined minimum, which is associated with the BPC formation. These results can be explained in terms of the relative stability of graphene, PG and PBC.

The total energy values per carbon atoms (in relation to graphene) of dehydrogenated PG and BPC are 1.26 and 0.63 eV, respectively. In this sense the removal of hydrogens from PG inverts the stability order in relation to BPC, BPC being now 0.63 eV per carbon atom more stable. Due to the similar topology the interconversion easily occurs, since it requires only ring rotations and the creation of new bonds forming the cyclobutadiene motif. These results strongly support that BPC can be obtained from PG by just selective dehydrogenation.

PG selective dehydrogenation is within our present-day synthesis capabilities. Recently, significant experimental advances have been produced in selective dehydrogenation of hydrocarbons \cite{otero,treier,liang,gutzler}. Fullerenes \cite{otero}, nanographene flakes \cite{treier}, and even two-dimensional networks \cite{gutzler,liang} similar to BPC have been already achieved. See more details in the supplementary materials. Obtaining BPC from DPG using these techniques is perfectly feasible. We hope the present results will stimulate further works along these lines.

\section{Acknowledgements}
This work was supported in part by the Brazilian Agencies CNPq, CAPES and FAPESP. 

\section{Supporting Information}

Brief discussion about simulations and MD videos. This information is available free of charge via Internet at http://pubs.acs.org


\newpage

\newpage
\clearpage

\end{document}